\documentstyle[multicol,aps]{revtex}

\def\baselinestretch{1.1}

\renewcommand{\narrowtext}{\begin{multicols}{2} \global\columnwidth20.5pc}
\renewcommand{\widetext}{\end{multicols} \global\columnwidth42.5pc}

\begin{document}

\title{\Large\bf  Restrictions on Gauge Groups in Noncommutative Gauge Theory }
\author{ \large\bf Keizo Matsubara\\[2mm]}

\address{\noindent   
Institute for Theoretical Physics, Uppsala University,
 Box 803, S-75108, Uppsala, Sweden\\
 keizo@teorfys.uu.se  \\[1mm] }

\date{March, 2000} 
\maketitle

\begin{abstract}
\noindent
We show that the gauge groups $SU(N)$, $SO(N)$ and $Sp(N)$ cannot be realized on a flat noncommutative manifold, while it is possible for $U(N)$.
\vspace{1mm}
\noindent
\end{abstract}

\narrowtext

\section{Introduction}
The Yang-Mills theories naturally arise as low energy limits of the 
theory of open strings. One can obtain Yang-Mills theories with
different gauge groups by studying different D-brane configurations
(see {\em e.g.} [1]).
For instance, if we place $N$ D-branes on top of each other in the flat space,
the corresponding open string theory gives rise to the Yang-Mills theory
with the gauge group $U(N)$. 
One can also obtain gauge theories with other gauge groups
such as $SO(N)$ and $Sp(N)$ by using the orientifold construction.
In more detail, one combines the spatial reflection $\sigma \rightarrow
\pi - \sigma $ on the string world-sheet with the target space reflection,
$X^\mu \rightarrow - X^\mu, \mu=1  ,\quad \dots, k$ and $X^\mu \rightarrow X^\mu, 
\mu=(k+1),\quad  \dots, 10$. It is the goal of this note to study which gauge groups
can be realized in the presence of the background $B$-field when
the brane world-volume turns into a noncommutative space [2,3,4].

Interaction with the $B$-field introduces an extra term into
the Polyakov action of the string [4],
\begin{equation}
\Delta S=\frac{-i}{2} \int_{\Sigma}B_{\mu \nu} \epsilon ^{ab} \partial_{a}X^{\mu}\partial_{b}X^{\nu}.
\end{equation} 
Here $a,b =1,2$ are world-sheet indices, $\Sigma$ is the worldsheet and $B_{\mu \nu} = - B_{\nu \mu}$
is the $B$-field on the target space. One requires the expression (1)
to be invariant with respect to the orientifold reflection.
This implies the following transformation rules for
components of the $B$-field,
\begin{eqnarray}
B_{\parallel \parallel} \rightarrow - B_{\parallel \parallel} & \qquad & 
B_{\perp \perp} \rightarrow - B_{\perp \perp} \nonumber \\
B_{\parallel \perp} \rightarrow B_{\parallel \perp} & \qquad & 
B_{\perp \parallel} \rightarrow B_{\perp \parallel}. 
\end{eqnarray}
Here the symbols  $\parallel$ and $\perp$ stand for the target space
indices $\mu=1, \dots, k$ and $\mu=(k+1), \dots, 10$, respectively.
In the orientifold construction we finally let the branes lie on the
orientifold.   The continuity of the $B$-field implies that 
the $B$-field on the brane, $B_{\parallel \parallel}$,
vanishes. Hence, the brane world-volumes are 
commutative since it is $B_{\parallel \parallel}$ which
is responsible for the noncommutativity [2,3,4].
This consideration indicates that one should encounter
difficulties in the construction of the gauge theories with
gauge groups $SO(N)$ and $Sp(N)$ on  noncommutative spaces.
Somewhat surprisingly, we find that the reduction from
$U(N)$ to $SU(N)$ in the framework of noncommutative
geometry also fails.

\section{The closure of classical Lie algebras under the Moyal commutator}
In the flat case the presence of a constant $B$-field
turns the D-branes into noncommutative spaces, with
the ordinary pointwise multiplication of functions replaced
by the Moyal product,
\begin{eqnarray}
(X * Y)(x) & = & \exp ( \frac{i}{2} \theta ^{ij} \partial^{x} _{i} \partial^{y} _{j})
X(x)Y(y)|_{x=y}  =  \nonumber \\
& = & XY + \frac{i}{2} \theta ^{ij} \partial _{i}X \partial _{j}Y  + ... 
\end{eqnarray}
Here $X$ and $Y$ are  functions on the D-brane world-volume,
and $\theta ^{ij}$ is a real-valued constant  antisymmetric tensor constructed
of the metric and $B$-field [3]. The Moyal product naturally
extends to $N$ by $N$ matrices, formula (3) still applies.
One can also introduce the Moyal commutator by the formula,
\begin{equation}
[X,Y]_{*} =X*Y - Y*X.
\end{equation}
In what follows we check whether the matrix Lie algebras
of the classical Lie groups $SO(N), U(N), SU(N)$ and $Sp(N)$ are
closed under the Moyal commutator.  We choose to work in
the fundamental representation of these Lie algebras.

The Lie algebra of $U(N)$ consists of anti-Hermitean matrices,
$\overline{X^t}=-X$, where the bar stands for
complex conjugation. We first show that this algebra
{\em is}  closed under the Moyal commutator.
The key observation is the following property
of the Moyal product,
\begin{equation}
\overline{(X*Y)^t}=\overline{Y^t} * \overline{X^t} .
\end{equation}
By using the ordinary rules for the transpose of matrices we get,
\begin{eqnarray}
& & (X * Y)^t = Y^t X^t + \nonumber \\
& & + \frac{i}{2} \theta ^{ij} \partial _{j}Y^t \partial _{i}X^t - \frac {1}{8}\theta ^{ij}\theta ^{kl}\partial _{j}\partial _{l}Y^t \partial _{i}\partial _{k}X^t + ... 
\end{eqnarray}
The construction for higher order terms is obvious. 
Now we apply the complex conjugation and rename the 
indices of $\theta$ to obtain,
\begin{eqnarray}
& &\overline{(X * Y)^t} =\overline{Y^t} \overline{X^t} + \frac{i}{2} \theta ^{ij}
\partial _{i}\overline{Y^t} \partial _{j}\overline{X^t} - \\
& &\frac {1}{8}\theta ^{ij}\theta ^{kl}\partial _{i}\partial _{k}\overline{Y^t}
\partial _{j}\partial _{l}\overline{X^t} + ... =\overline{Y^t}*\overline{X^t}.
\end{eqnarray}
Taking into account $\overline{X^t}=-X$ and $\overline{Y^t}=-Y$ yields,
\begin{eqnarray}
\overline{[X,Y]_{*}^{t}} & = & \overline{(X*Y)^t} -\overline{Y*X)^t}= \\
& = & \overline{Y^t}*\overline{X^t} - \overline{X^t}*\overline{Y^t} = \\ 
& = & Y*X - X*Y =-[X,Y]_{*}
\end{eqnarray}
which shows that the algebra $U(N)$ is closed under the Moyal commutator.

We now turn to the algebras of $SO(N)$, $SU(N)$ and $Sp(N)$. 
We first show that for $N=2$ these algebras  are not closed
with respect to the Moyal commutator. 

The counter examples for both $SO(2)$ and $Sp(2)$ are  given
by formulas,
\begin{equation}
X=
\left(
\begin{array}{cc}
0 & \alpha \\
-\alpha & 0  \\
\end{array} 
\right)
\qquad
Y=
\left(
\begin{array}{cc}
0 & \beta \\
-\beta & 0  \\
\end{array} 
\right).
\end{equation}
and the counterexample for $SU(2)$ is
\begin{equation}
X=
\left(
\begin{array}{cc}
i\alpha & 0 \\
0 & -i\alpha  \\
\end{array} 
\right)
Y=
\left(
\begin{array}{cc}
i\beta & 0 \\
0 & -i\beta  \\
\end{array} 
\right).
\end{equation}
Here $\alpha$ and $\beta$ are coordinates on the manifold 
chosen so that $\theta ^{\alpha \beta} \neq 0 $.
This can always be done unless $\theta=0$
and the Moyal product coincides with the ordinary
multiplication of matrix-valued functions.
With $X$ and $Y$ as given above
one can easily compute the Moyal commutator
since all derivatives of order higher than one vanish. 
The result for both counter examples is
\begin{equation}
[X,Y]_{*}= i\theta ^{\alpha \beta}
\left(
\begin{array}{cc}
-1 & 0 \\
0 & -1  \\
\end{array} 
\right) .
\end{equation}
Note that this matrix has a nonvanishing trace.
Since the Lie algebras of $SO(2), SU(2)$ and $Sp(2)$
consist of traceless matrices, we conclude that
they are not closed under the Moyal commutator.
This also applies to $SO(N), SU(N)$ and $Sp(N)$ for
arbitrary $N$ because they contain $SO(2), SU(2)$ and $Sp(2)$
as subgroups.

{\bf Acknowledgements}
First of all I would like to thank Anton Alekseev who has been very helpful during the preparation of this note. I would also like to thank Ulf Danielsson, Martin Kruczenski and M\aa rten Stenmark for helpful and encouraging discussions.

\def\baselinestretch{1.0}

\widetext
\end{document}